# Model nuclear energy density functionals derived from *ab initio* calculations

**G. Salvioni**[1,2], **J. Dobaczewski**[1,2,3,4], **C. Barbieri**[5], **G. Carlsson**[6], **A. Idini** [5,6], **A. Pastore**[3]

[1]Department of Physics, P.O. Box 35 (YFL), FI-40014 University of Jyväskylä, Finland
[2]Helsinki Institute of Physics, P.O. Box 64, FI-00014University of Helsinki, Finland
[3]Department of Physics, University of York, Heslington, York YO10 5DD, United Kingdom
[4]Institute of Theoretical Physics, Faculty of Physics, University of Warsaw, ul. Pasteura 5, PL-02-093 Warsaw, Poland
[5]Department of Physics, University of Surrey, Guildford, GU2 7XH, United Kingdom
[6]Division of Mathematical Physics, Department of Physics, LTH, Lund University, P.O. Box 118, S-22100 Lund, Sweden

E-mail: `gianlucasalvioni@gmail.com`

**Abstract.** We present the first application of a new approach, proposed in [Journal of Physics G: Nuclear and Particle Physics, vol. 43, 04LT01 (2016)] to derive coupling constants of the Skyrme energy density functional (EDF) from *ab initio* Hamiltonian. By perturbing the *ab initio* Hamiltonian with several functional generators defining the Skyrme EDF, we create a set of metadata that is then used to constrain the coupling constants of the functional. We use statistical analysis to obtain such an *ab initio*-equivalent Skyrme EDF. We find that the resulting functional describes properties of atomic nuclei and infinite nuclear matter quite poorly. This may point out to the necessity of building up the *ab initio*-equivalent functionals from more sophisticated generators. However, we also indicate that the current precision of the *ab initio* calculations may be insufficient for deriving meaningful nuclear EDFs.

## 1. Introduction

Different energy scales that appear in nuclear systems suggest a theoretical approach based on effective field theories (EFT), which use relevant degrees of freedom adapted to a given energy scale [1]. A remarkable example is the chiral effective field theory ($\chi$-EFT) [2, 3]: by using neutrons, protons, and pions as degrees of freedom, $\chi$-EFT is able to provide consistent description of numerous observables in atomic nuclei. Chiral interactions are used in the framework of *ab initio* methods to solve the many-body

Schrödinger equations, employing controlled approximations. The numerical solutions require huge computational resources; for this reason, actual state-of-the-art *ab initio* calculations are limited to nuclei that are not too heavy and/or close to (semi)magic systems [4, 5, 6, 7]. In particular, the description of open-shells nuclei has a recent history, started with the Gorkov-Green's Function approach [8].

Another example of successful effective theory is the approach based on nuclear energy density functionals (EDFs) [9, 10]. Nuclear EDF is a versatile tool that, at low computational cost, allows us to describe properties of atomic nuclei across the entire nuclear chart from drip-line to drip-line and from light to super-heavy nuclei. The underlying non-relativistic functionals are usually obtained from phenomenological two-body potentials called functional generators, which have radial form factors of zero-range, for Skyrme [11, 12], or finite-range, for Gogny [13] implementations, with coupling constants adjusted to reproduce selected nuclear observables. In addition to finite nuclei, also infinite nuclear matter properties can be addressed within the EDFs formalism [14, 15, 16], and included among the observables.

In Ref. [17], it was shown that Skyrme EDFs have reached their best in terms of reproducing the experimental observables. As a consequence, several groups are exploring new forms of functional generators to improve precision of describing data [18, 19, 20, 21, 22]. The resulting functionals are then adjusted following standard procedures based on constraining the coupling constants to a given set of nuclear observables [23, 24].

In this article, we aim to explore a complementary approach, which is based on explicitly bridging the *ab initio* methods with the nuclear EDFs. Formal connections between the two approaches are the subject of recent studies [25, 26]. Furthermore, in Ref. [27] the density matrix expansion formalism has been used to obtain a set of functionals microscopically constrained from chiral effective field theory interactions. Here, we proceed by fitting parameters that enter the energy functional directly to metadata generated by the *ab initio* calculations, using the method suggested in Ref. [28]. In Ref. [28] this method was applied adjusting the Skyrme coupling constants to metadata generated by the Gogny functional. In this work, we perform a realistic application of the method by employing full-fledged *ab initio* calculations that use Self-Consistent Green's Function (SCGF) [29] methodology with the chiral interaction NNLO$_{\text{sat}}$ [30].

The paper is organised as follows: Section 2 briefly recalls the SCGF (2.1) and EDF (2.2) methods, and explains the formalism to derive the model functionals (2.3). Results are discussed in Section 3 and conclusions are given in Section 4.

## 2. Theoretical

### 2.1. Self-Consistent Green's Function method

We separate the nuclear many-body Hamiltonian $\hat{H} = \hat{H}_0 + \hat{H}_1$ into a noninteracting part $\hat{H}_0 = \hat{T} + \hat{U}$, with the auxiliary one-body operator $\hat{U}$, and the interacting part defined as $\hat{H}_1 = -\hat{U} + \hat{V}^{2\mathrm{B}} + \hat{V}^{3\mathrm{B}}$. We express the nuclear Hamiltonian in second quantisation as

$$\hat{H} = \sum_{\alpha} \epsilon_{\alpha}^0 a_{\alpha}^{\dagger} a_{\alpha} - \sum_{\alpha\beta} \langle \alpha | \hat{U} | \beta \rangle a_{\alpha}^{\dagger} a_{\beta} + \frac{1}{4} \sum_{\alpha\beta\gamma\delta} \langle \alpha\beta | \hat{V}^{2\mathrm{B}} | \gamma\delta \rangle a_{\alpha}^{\dagger} a_{\beta}^{\dagger} a_{\delta} a_{\gamma}$$
$$+ \frac{1}{36} \sum_{\alpha\beta\mu\gamma\delta\nu} \langle \alpha\beta\mu | \hat{V}^{3\mathrm{B}} | \gamma\delta\nu \rangle a_{\alpha}^{\dagger} a_{\beta}^{\dagger} a_{\mu}^{\dagger} a_{\nu} a_{\delta} a_{\gamma}, \tag{1}$$

where $\epsilon_{\alpha}^0$ are the single-particle energies of $\hat{H}_0$. By solving the corresponding Schrödinger equation,

$$\hat{H} | \Psi_0^A \rangle = E_0^A | \Psi_0^A \rangle, \tag{2}$$

one obtains the ground-state energy $E_0^A$ and wave function $| \Psi_0^A \rangle$ of the nuclear system. The exact solution of Eq. (2) is very complicated and it is typically limited to systems with very few number of nucleons [31]. Rather than calculating the full many-body wave function, it is possible to expand the solution of the Schrödinger equation in terms of the propagation of single-particle excitations and the correlated density matrix of the system by using SCGF method [29]. These excitations represent basic collective degrees of freedom of the nucleus and they are described by the propagator of one-body Green's function $G_{\alpha\beta}$. Using the Källén-Lehmann representation [32, 33], $G_{\alpha\beta}$ takes the form

$$G_{\alpha\beta}(E) = \sum_n \frac{\langle \Psi_0^A | a_{\alpha} | \Psi_n^{A+1} \rangle \langle \Psi_n^{A+1} | a_{\beta}^{\dagger} | \Psi_0^A \rangle}{E - (E_n^{A+1} - E_0^A) + i\eta}$$
$$+ \sum_k \frac{\langle \Psi_0^A | a_{\beta}^{\dagger} | \Psi_k^{A-1} \rangle \langle \Psi_k^{A-1} | a_{\alpha} | \Psi_0^A \rangle}{E - (E_0^A - E_k^{A-1}) - i\eta}. \tag{3}$$

Here the complete set of eigenstates $| \Psi_n^{A+1} \rangle$, $| \Psi_k^{A-1} \rangle$ with eigenvalues $E_n^{A+1}$, $E_k^{A-1}$ introduce the intermediate $(A \pm 1)$-body systems. Then, $E_n^{A+1} - E_0^A$ and $E_0^A - E_k^{A-1}$ are respectively the excitation energies of the propagating quasi-particle (index $n$) and quasi-hole (index $k$) states.

The propagator given in Eq. (3) satisfies the Dyson equation

$$G_{\alpha\beta}(E) = G_{\alpha\beta}^{(0)}(E) + \sum_{\gamma\delta} G_{\alpha\gamma}^{(0)}(E) \, \Sigma_{\gamma\delta}^{\star}(E) \, G_{\delta\beta}(E), \tag{4}$$

where $G_{\alpha\beta}^{(0)}(E)$ is the propagator for the noninteracting system $\hat{H}_0$, and $\Sigma_{\gamma\delta}^{\star}(E)$ is the irreducible self-energy. The latter represents the nonlocal and energy-dependent potential to which each nucleon is subjected when interacting within the nuclear medium. The nonlinearity of Eq. (4) in terms of $G_{\alpha\beta}(E)$ requires an iterative procedure to reach convergence. The full self-consistency is required to satisfy fundamental symmetries and conservation laws [34].

A crucial ingredient of Eq. (4) is the self-energy. This is composed of three parts as

$$\Sigma_{\gamma\delta}^{\star}(E) = -\langle \gamma|U|\delta \rangle + \Sigma_{\gamma\delta}^{(\infty)} + \tilde{\Sigma}_{\gamma\delta}(E), \tag{5}$$

which, respectively, are the auxiliary potential, static mean-field, and energy-dependent component. To perform calculations of the energy-dependent component, in this work we employ the Algebraic Diagrammatic Construction method [35, 36] up to third order, denoted by ADC(3). We refer to Ref. [37] for a more pedagogical introduction of the method. The accuracy of ADC(3) can be estimated to be of the fourth order in $\hat{H}_1$, giving in practice an error of 1% of the total binding energy [37].

Another important aspect of the *ab initio* calculations is the presence of explicit three-body interaction $\hat{V}^{3B}$. By means of the modified Migdal-Galitski-Koltun sum rule [38], we can write the ground-state energy of the system as,

$$E_0^A = \sum_{\alpha\beta} \frac{1}{2\pi} \int_{-\infty}^{\epsilon_0^-} dE \left[ \langle \alpha|\hat{T}|\beta \rangle + E\,\delta_{\alpha\beta} \right] \text{Im}\{G_{\beta\alpha}(E)\} - \frac{1}{2} \langle \Psi_0^A|\hat{V}^{3B}|\Psi_0^A \rangle, \tag{6}$$

where $\epsilon_0^-$ is the highest quasi-hole energy, namely $\epsilon_0^- = \max_k(E_0^A - E_k^{A-1})$. The determination of the expectation value of the three-body interaction would require calculation of many-body propagators. Instead, assuming that the magnitude of the three-body term is smaller than that of the two-body term, we take the lowest order approximation for the Hamiltonian, leading to

$$\langle \Psi_0^A|\hat{V}^{3B}|\Psi_0^A \rangle \approx \frac{1}{6} \sum_{\alpha\beta\mu\gamma\delta\nu} \langle \alpha\beta\mu|\hat{V}^{3B}|\gamma\delta\nu \rangle \; \rho_{\gamma\alpha}\rho_{\delta\beta}\rho_{\nu\mu}, \tag{7}$$

where $\rho_{\alpha\beta}$ is the one-body density matrix.

## 2.2. Model Energy Density Functionals and generators

Having defined the main theoretical aspects of the *ab initio* method employed in this article, we now define the model energy density functional as

$$\tilde{E}[\rho] = T^{1B}[\rho] + V^{\text{Coul}}[\rho] + \sum_j C_j \; V_j^{\text{gen}}[\rho] \; . \tag{8}$$

The different terms correspond to average values, evaluated with respect to a Hartree-Fock (HF) state, of the kinetic energy $T^{1B}[\rho] \equiv \langle \Phi|\hat{T}^{1B}|\Phi \rangle_{\text{HF}}$, Coulomb potential $V^{\text{Coul}}[\rho] \equiv \langle \Phi|\hat{V}^{\text{Coul}}|\Phi \rangle_{\text{HF}}$, and interaction components $V_j^{\text{gen}}[\rho] \equiv \langle \Phi|\hat{V}_j^{\text{gen}}|\Phi \rangle_{\text{HF}}$. The operators $\hat{V}_j^{\text{gen}}$ represent our choice of the generators to build the functional, whereas $C_j$ are the coupling constants we need to adjust on (meta)-data.

In the present article, we define generators $\hat{V}_j^{\text{gen}}$, based on ten individual terms $\hat{T}_i$ and $\hat{T}_i^{\sigma}$ for $i = 0, 1, 2$, and $\hat{T}_e$, $\hat{T}_o$, $\hat{T}_{W0}$, and $\hat{T}_3$ of the Skyrme functional generator [12], that is,

$$
\begin{aligned}
\hat{V}^{\text{Skyrme}} = {} & t_0(1 + x_0\hat{P}^{\sigma})\delta(\boldsymbol{r_1} - \boldsymbol{r_2}) \\
& + \frac{1}{2}t_1(1 + x_1\hat{P}^{\sigma})\left[\hat{\boldsymbol{k}}'^2\delta(\boldsymbol{r_1} - \boldsymbol{r_2}) + \delta(\boldsymbol{r_1} - \boldsymbol{r_2})\hat{\boldsymbol{k}}^2\right] \\
& + t_2(1 + x_2\hat{P}^{\sigma})\hat{\boldsymbol{k}}' \cdot \delta(\boldsymbol{r_1} - \boldsymbol{r_2})\hat{\boldsymbol{k}} \\
& + iW_0(\hat{\boldsymbol{\sigma_1}} + \hat{\boldsymbol{\sigma_2}}) \cdot \left[\hat{\boldsymbol{k}}' \times \delta(\boldsymbol{r_1} - \boldsymbol{r_2})\hat{\boldsymbol{k}}\right] \\
& + \frac{t_e}{2}\Big\{\left[3(\boldsymbol{\sigma_1} \cdot \boldsymbol{k}')(\boldsymbol{\sigma_2} \cdot \boldsymbol{k}') - (\boldsymbol{\sigma_1} \cdot \boldsymbol{\sigma_2})\boldsymbol{k}'^2\right]\delta(\boldsymbol{r_1} - \boldsymbol{r_2}) \\
& + \delta(\boldsymbol{r_1} - \boldsymbol{r_2})\left[3(\boldsymbol{\sigma_1} \cdot \boldsymbol{k})(\boldsymbol{\sigma_2} \cdot \boldsymbol{k}) - (\boldsymbol{\sigma_1} \cdot \boldsymbol{\sigma_2})\boldsymbol{k}^2\right]\Big\} \\
& + t_o\Big\{3(\boldsymbol{\sigma_1} \cdot \boldsymbol{k}')\delta(\boldsymbol{r_1} - \boldsymbol{r_2})(\boldsymbol{\sigma_2} \cdot \boldsymbol{k}) - (\boldsymbol{\sigma_1} \cdot \boldsymbol{\sigma_2})\left[\boldsymbol{k}' \cdot \delta(\boldsymbol{r_1} - \boldsymbol{r_2})\boldsymbol{k}\right]\Big\} \\
& + t_3\delta(\boldsymbol{r_1} - \boldsymbol{r_2})\delta(\boldsymbol{r_2} - \boldsymbol{r_3})\mathcal{A}_{123}(\hat{P}^{\sigma}, \hat{P}^{\tau}) \\
= {} & t_0\hat{T}_0 + t_0x_0\hat{T}_0^{\sigma} + t_1\hat{T}_1 + t_1x_1\hat{T}_1^{\sigma} + t_2\hat{T}_2 + t_2x_2\hat{T}_2^{\sigma} \\
& + W_0\hat{T}_{W0} + t_e\hat{T}_e + t_o\hat{T}_o + t_3\hat{T}_3,
\end{aligned}
\tag{9}
$$

where $\mathcal{A}_{123}$ is the antisymmetric operator for combinations of $\hat{P}^{\sigma} = \frac{1}{2}(1 + \boldsymbol{\sigma_1} \cdot \boldsymbol{\sigma_2})$ and $\hat{P}^{\tau} = \frac{1}{2}(1 + \boldsymbol{\tau_1} \cdot \boldsymbol{\tau_2})$. We refer to Ref. [39] for more details.

To determine the coupling constants of the interaction using *ab initio* methods, it is convenient to transform the generators given in Eqs. (9) to linear combinations that give specific isoscalar/isovector terms of the functional [28]. This is done by means of the following matrix relation,

$$
\begin{bmatrix}
\hat{V}_0^{\rho} \\
\hat{V}_1^{\rho} \\
\hat{V}_0^{\Delta\rho} \\
\hat{V}_1^{\Delta\rho} \\
\hat{V}_0^{\tau} \\
\hat{V}_1^{\tau} \\
\hat{V}_0^{J1} \\
\hat{V}_1^{J1} \\
\hat{V}_{W0} \\
\hat{V}_{t3}
\end{bmatrix}
=
\begin{bmatrix}
\frac{8}{3} & -\frac{4}{3} & 0 & 0 & 0 & 0 & 0 & 0 & 0 & 0 \\
0 & -4 & 0 & 0 & 0 & 0 & 0 & 0 & 0 & 0 \\
0 & 0 & -\frac{16}{3} & \frac{8}{3} & \frac{16}{3} & -\frac{8}{3} & -\frac{16}{15} & \frac{16}{15} & 0 & 0 \\
0 & 0 & 0 & 8 & -\frac{32}{3} & \frac{40}{3} & \frac{16}{5} & \frac{16}{15} & 0 & 0 \\
0 & 0 & \frac{4}{3} & -\frac{2}{3} & 4 & -2 & -\frac{8}{15} & 0 & 0 & 0 \\
0 & 0 & 0 & -2 & -8 & 10 & \frac{8}{5} & 0 & 0 & 0 \\
0 & 0 & 0 & 0 & 0 & 0 & \frac{8}{5} & \frac{8}{5} & 0 & 0 \\
0 & 0 & 0 & 0 & 0 & 0 & -\frac{24}{5} & \frac{8}{5} & 0 & 0 \\
0 & 0 & 0 & 0 & 0 & 0 & 0 & 0 & 1 & 0 \\
0 & 0 & 0 & 0 & 0 & 0 & 0 & 0 & 0 & 1
\end{bmatrix}
\begin{bmatrix}
\hat{T}_0 \\
\hat{T}_0^{\sigma} \\
\hat{T}_1 \\
\hat{T}_1^{\sigma} \\
\hat{T}_2 \\
\hat{T}_2^{\sigma} \\
\hat{T}_e \\
\hat{T}_o \\
\hat{T}_{W0} \\
\hat{T}_3
\end{bmatrix}.
\tag{10}
$$

We explicitly included only the vector part $\hat{V}_T^{J1}$ of the tensor term [40], which in the case of spherical symmetry considered here gives the only non-zero contribution. Note that the terms associated with the $\hat{V}_{W0}$ and $\hat{V}_{t3}$ generators do not allow for the separation of isoscalar/isovector terms, and therefore we keep them identical to the corresponding terms of the Skyrme generator in Eq. (9). The generators listed on the left-hand side of Eq. (10) are then used as generators, $\hat{V}_j^{\text{gen}}$, $j = 1 - 10$, that define the functional in Eq. (8).

## 2.3. Derivation of the functionals

In the electronic density functional theory [41], one uses the Levy-Lieb constrained variation [42, 43, 44, 45] to obtain the ground state energy of the system. This procedure consists in a two-step minimisation,

$$E_{\text{g.s.}} = \min_{\rho} \left\{ \min_{|\Psi\rangle \to \rho} \left[ \langle \Psi | \hat{T} + \hat{V} | \Psi \rangle \right] \right\} \equiv \min_{\rho} E[\rho], \tag{11}$$

where $\hat{V}$ stands for the Coulomb potential and symbol $\min_{|\Psi\rangle \to \rho}$ denotes an inner (first-step) minimisation over all correlated many-body states $|\Psi\rangle$ that have a common fixed one-body density profile $\rho(\boldsymbol{r})$. The outer (second-step) minimisation is then performed over all possible profiles $\rho(\boldsymbol{r})$, and, in such a way, the global minimum of energy, and thus the exact ground-state energy $E_{\text{g.s.}}$ is obtained.

The inner minimisation can be conveniently performed by an unconstrained minimisation of the Routhian $\hat{R}$ at fixed one-body external potential $U(\boldsymbol{r})$ that plays a role of the Lagrange multiplier,

$$R[U] = \min_{|\Psi\rangle} \langle \Psi | \hat{R} | \Psi \rangle = \min_{|\Psi\rangle} \langle \Psi | \left[ \hat{T} + \hat{V} + \int \mathrm{d}\boldsymbol{r} \ U(\boldsymbol{r}) \hat{\rho}(\boldsymbol{r}) \right] | \Psi \rangle. \tag{12}$$

This gives the energy $E[U] = R[U] - \int \mathrm{d}\boldsymbol{r} \ U(\boldsymbol{r})\rho(\boldsymbol{r})$ and density $\rho[U]$ as functionals of the potential $U(\boldsymbol{r})$. Assuming that the inverse functional $U[\rho]$ can be found, we obtain the exact energy density functional,

$$E[\rho] \equiv E[U[\rho]], \tag{13}$$

which after the outer (second-step) minimisation gives again the exact ground-state energy $E_{\text{g.s.}}$.

In the case of a nuclear system, we consider the analogous first-step variation consisting in the minimisation of the Routhian $\hat{R}^{\text{ab}}$ as

$$\delta_{\Psi} \langle \Psi | \hat{R}^{\text{ab}} | \Psi \rangle = \delta_{\Psi} \langle \Psi | \left[ \hat{H}^{\text{ab}} + \int \mathrm{d}\boldsymbol{r} \ U(\boldsymbol{r}) \hat{\rho}(\boldsymbol{r}) \right] | \Psi \rangle = 0, \tag{14}$$

where $\hat{H}^{\text{ab}}$ is the Hamiltonian of the system, Eq. (1). We use the superscript ab to indicate that it is the Hamiltonian of the *ab initio* theory, distinguishing it from the Hamiltonian used to build the model functional. The minimisation gives us many-body states as functionals of the external potential, $|\Psi(U)\rangle$.

The integrand on the right-hand side of Eq. (14) introduces a perturbation of the ground-state $|\Psi_{\text{g.s.}}\rangle$. The response of the system to the perturbation causes a change in the density $\rho$. If we were able to probe the system with all possible potentials $U(\boldsymbol{r})$, we would have obtained the functional $E^{\text{ab}}[U] \equiv \langle \Psi(U) | \hat{H}^{\text{ab}} | \Psi(U) \rangle$, and then, as above, the exact energy density functional $E^{\text{ab}}[\rho]$. In the second step, the functional $E^{\text{ab}}[\rho]$ is

minimised with respect to $\rho$, which gives the exact ground-state energy $E^{\mathrm{ab}}_{\mathrm{g.s.}}$ and density $\rho_{\mathrm{g.s.}}(\boldsymbol{r})$.

Being unable to perturb the system with an infinite number of external potentials, let us introduce a discrete finite set of pre-defined external potentials $u_i(\boldsymbol{r})$ and their corresponding strengths $\lambda_i$, whereby the first-step minimisation (14) becomes,

$$\delta_\Psi \langle \Psi | \hat{R}^{\mathrm{ab}} | \Psi \rangle = \delta_\Psi \langle \Psi | \left[ \hat{H}^{\mathrm{ab}} + \sum_i \lambda_i \int \mathrm{d}\boldsymbol{r}\, u_i(\boldsymbol{r}) \hat{\rho}(\boldsymbol{r}) \right] | \Psi \rangle = 0. \qquad (15)$$

With this restriction, the *ab initio* energy and density, $E^{\mathrm{ab}}(\lambda_i)$ and $\rho^{\mathrm{ab}}(\lambda_i)$, become functions (not functionals) of the finite set of strengths $\lambda_i$. Obviously, we now do not know what is the full energy density functional $E^{\mathrm{ab}}[\rho]$ in the infinite-dimensional space of all possible one-body density profiles, however, we know it exactly on a finite-dimensional manifold of densities parametrised by strengths $\lambda_i$. Recall that this manifold still contains the point $\lambda_i = 0$ that corresponds to the exact ground state. The second-step variation would now correspond to the minimisation of function $E^{\mathrm{ab}}(\lambda_i)$ in finite dimensions.

As proposed in Ref. [28], we now conjecture that a meaningful manifold of *ab initio* densities $\rho^{\mathrm{ab}}(\lambda_i)$ can be obtained not by pre-defining external potentials $u_i(\boldsymbol{r})$, but by perturbing the system with generators that are going to be used for modelling the functional, Sec. 2.2, that is,

$$\delta_\Psi \langle \Psi | \hat{R}^{\mathrm{ab}} | \Psi \rangle = \delta_\Psi \langle \Psi | \left[ \hat{H}^{\mathrm{ab}} + \sum_i \lambda_i \hat{V}^{\mathrm{gen}}_i \right] | \Psi \rangle = 0. \qquad (16)$$

Since the unrestricted minimisation of the Routhian is equivalent to finding its exact ground state $|\Psi^{\mathrm{ab}}(\lambda_i)\rangle$ and eigenvalue $R^{\mathrm{ab}}(\lambda_i)$, we have

$$\left[ \hat{H}^{\mathrm{ab}} + \sum_i \lambda_i \hat{V}^{\mathrm{gen}}_i \right] | \Psi^{\mathrm{ab}}(\lambda_i) \rangle = R^{\mathrm{ab}}(\lambda_i) | \Psi^{\mathrm{ab}}(\lambda_i) \rangle. \qquad (17)$$

Formally, by replacing minimisation (15) with minimisation (16), we do not make any new assumption. Indeed, the previously made assumption about the reversibility of the functional $\rho[U]$ suffices. It stipulates that for any density $\rho(\boldsymbol{r})$ generated by the latter minimisation there exists a potential $U(\boldsymbol{r})$ that would have generated it by the former minimisation. However, when using minimisation (16) we do not have to pre-define any potential or, for that matter, we do not have to know it at all.

Since our goal is to use the *ab initio* energies with perturbation $\lambda_i$ as metadata to determine the coupling constants of our model functional, we impose that the *ab initio* energy can be expressed in the form of functional (8), that is,

$$E^{\mathrm{ab}}(\lambda_i) = T^{\mathrm{1B}}[\rho^{\mathrm{ab}}(\lambda_i)] + V^{\mathrm{Coul}}[\rho^{\mathrm{ab}}(\lambda_i)] + \sum_j C_j\, V^{\mathrm{gen}}_j[\rho^{\mathrm{ab}}(\lambda_i)]\,. \qquad (18)$$

Considering that the Kohn-Sham kinetic energy represents a good approximation to the one-body kinetic energy, we assume that

$$\langle \Psi^{\mathrm{ab}}(\lambda_i)|\hat{T}^{\mathrm{1B}}|\Psi^{\mathrm{ab}}(\lambda_i)\rangle \approx T^{\mathrm{1B}}[\rho^{\mathrm{ab}}(\lambda_i)]. \tag{19}$$

If in the *ab initio* Hamiltonian the two-body center-of-mass correction $\hat{T}^{\mathrm{2B}}$ is included, it needs to be removed from the left-hand side of Eq. (18). Furthermore, we suppose that the Coulomb contribution in the *ab initio* Hamiltonian is close to the Hartree-Fock average in the functional, namely,

$$\langle \Psi^{\mathrm{ab}}(\lambda_i)|\hat{V}^{\mathrm{Coul}}|\Psi^{\mathrm{ab}}(\lambda_i)\rangle \approx V^{\mathrm{Coul}}[\rho^{\mathrm{ab}}(\lambda_i)]. \tag{20}$$

With these two additional assumptions, we subtract the kinetic and the Coulomb energies from both sides of Eq. (18), rewriting it as

$$\langle \Psi^{\mathrm{ab}}(\lambda_i)|\hat{V}^{\mathrm{ab}}|\Psi^{\mathrm{ab}}(\lambda_i)\rangle = \langle \Psi^{\mathrm{ab}}(\lambda_i)|\hat{H}^{\mathrm{ab}} - \hat{T} - \hat{V}^{\mathrm{Coul}}|\Psi^{\mathrm{ab}}(\lambda_i)\rangle = \sum_j C_j \, V_j^{\mathrm{gen}}[\rho^{\mathrm{ab}}(\lambda_i)], \tag{21}$$

where $\hat{V}^{\mathrm{ab}}$ is the *ab initio* potential. The coupling constants $C_j$ can now be obtained by linear regression analysis to best match the *ab initio* results appearing on both sides of Eq. (21) and determined for a meaningful set the strength parameters $\lambda_i$.

To evaluate the right-hand side of Eq. (21), we used standard expressions for Hartree-Fock expectation values of two- and three-body generators, that is,

$$\langle \Psi(\lambda_i)|\hat{V}_{\mathrm{2B}}^{\mathrm{gen}}|\Psi(\lambda_i)\rangle_{\mathrm{HF}} = \frac{1}{2} \sum_{\alpha\beta\gamma\delta} \langle \alpha\beta|\hat{V}_{\mathrm{2B}}^{\mathrm{gen}}|\gamma\delta\rangle \rho_{\gamma\alpha}^{\mathrm{ab}}(\lambda_i)\rho_{\delta\beta}^{\mathrm{ab}}(\lambda_i), \tag{22}$$

$$\langle \Psi(\lambda_i)|\hat{V}_{\mathrm{3B}}^{\mathrm{gen}}|\Psi(\lambda_i)\rangle_{\mathrm{HF}} = \frac{1}{6} \sum_{\alpha\beta\mu\gamma\delta\nu} \langle \alpha\beta\mu|\hat{V}_{\mathrm{3B}}^{\mathrm{gen}}|\gamma\delta\nu\rangle \, \rho_{\gamma\alpha}^{\mathrm{ab}}(\lambda_i)\rho_{\delta\beta}^{\mathrm{ab}}(\lambda_i)\rho_{\nu\mu}^{\mathrm{ab}}(\lambda_i), \tag{23}$$

where $\rho_{\alpha\beta}^{\mathrm{ab}}$ is the *ab initio* one-body density matrix, and $\langle \alpha\beta|\hat{V}_{\mathrm{2B}}^{\mathrm{gen}}|\gamma\delta\rangle$ and $\langle \alpha\beta\mu|\hat{V}_{\mathrm{3B}}^{\mathrm{gen}}|\gamma\delta\nu\rangle$ are the antisymmetrized two- and three-body matrix elements.

To evaluate the left-hand side of Eq. (17), we have to take into account the fact that the SCGF solver is not able to separate the different potential contributions to the Routhian in Eq. (16), but it provides us with the total interaction energy, defined as

$$V^{\mathrm{tot}}(\lambda_i) = \langle \Psi^{\mathrm{ab}}(\lambda_i)|\hat{V}^{\mathrm{ab}} + \hat{V}^{\mathrm{Coul}} + \sum_i \lambda_i \hat{V}_i^{\mathrm{gen}}|\Psi^{\mathrm{ab}}(\lambda_i)\rangle. \tag{24}$$

This gives Eq. (21) in the form

$$V^{\mathrm{ab}}(\lambda_i) \equiv V^{\mathrm{tot}}(\lambda_i) - \langle \Psi^{\mathrm{ab}}(\lambda_i)|\hat{V}^{\mathrm{Coul}} + \sum_i \lambda_i \hat{V}_i^{\mathrm{gen}}|\Psi^{\mathrm{ab}}(\lambda_i)\rangle = \sum_j C_j \, V_j^{\mathrm{gen}}[\rho^{\mathrm{ab}}(\lambda_i)], \tag{25}$$

that is, we have to evaluate the *ab initio* expectation values of the Coulomb potential and functional generators.

For a generic two-body operator, the exact expectation value $\langle \hat{O}^{\mathrm{2B}} \rangle \equiv \langle \Psi(\lambda_i)|\hat{O}^{\mathrm{2B}}|\Psi(\lambda_i)\rangle$ is given as infinite expansion in terms of the effective interaction and

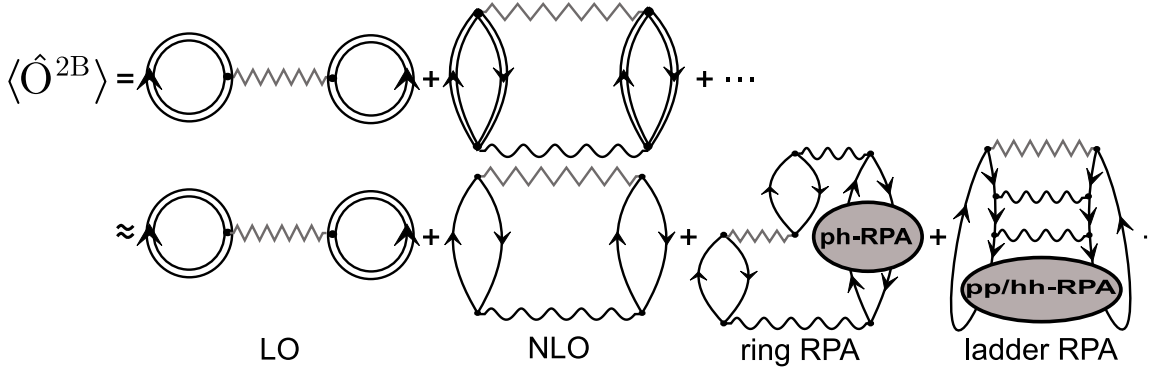

**Figure 1.** Diagrammatic representation of the expectation value of the two-body operator $\hat{O}^{2B}$, represented by zigzag lines. Double straight lines are used for dressed propagators, single lines for OpRS propagators, and wavy lines for two-body effective interactions. These are Feynman diagrams in the energy formulation, that is, they include forward and backward propagation. In the right-hand side of the first line, we can recognise the infinite expansion in term of dressed propagators. In the second line, we use the $ph$-RPA and $pp/hh$-RPA insertions to estimate contributions from NNLO and higher order terms.

dressed propagators, as sketched in Fig. 1 (first line). From a practical point of view, such a summation becomes computationally difficult, because the dressed propagators contain many poles that multiply matrix elements of every interaction line. Therefore, we approximate the single-particle propagator $G_{\alpha\beta}(E)$ with an optimised reference state (OpRS) propagator [46, 47], defined as

$$G_{\alpha\beta}^{OpRS}(E) = \sum_{n \notin F} \frac{(\phi_\alpha^n)^* \phi_\beta^n}{E - \epsilon_n^{OpRS} + i\eta} + \sum_{k \in F} \frac{\phi_\alpha^k (\phi_\beta^k)^*}{E - \epsilon_k^{OpRS} - i\eta}. \tag{26}$$

This is a model propagator for independent-particle states with energy $\epsilon^{OpRS}$ and wave function $\phi$. Such a propagator contains a reduced number of poles compared to the dressed one, while the energies and wave functions of the OpRS propagator are constrained to give the same momenta of the dressed propagators. We estimate $\langle\hat{O}^{2B}\rangle$ using the dressed propagators in the leading order (LO), or Hartree-Fock average, and the OpRS propagators in the next-to-leading order (NLO) and in the all-orders resummation diagrams, as depicted in the second line of Fig. 1. We account for higher orders with the $ph$-, $pp$- and $hh$-RPA (Random Phase Approximation) insertions respectively in the ring and ladder resummation diagrams. At the cost of introducing a small error of the density, the use of the OpRS propagator allows us to remarkably speed up calculation of $\langle\hat{V}^{Coul}\rangle$ and $\langle\hat{V}_i^{gen}\rangle$, which, otherwise, would not have been possible.

## 3. Results

In this section, we present our results obtained using SCGF with ADC(3) approximation. We chose the interaction NNLO$_{sat}$ [30] for the two-body and three-body sector. We made this choice since this interaction has been optimised to reproduce ground state energies and radii for isotopes up to mass $A = 24$ and it has been shown to predict accurate saturation properties also for larger isotopes and for infinite nuclear matter [48, 49, 50, 51].

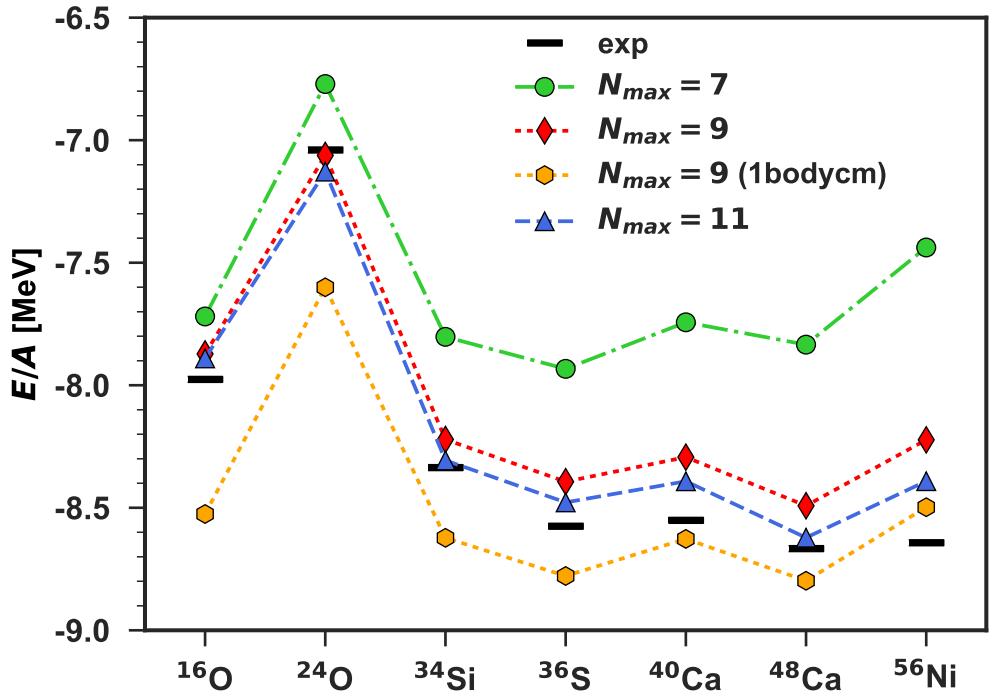

**Figure 2.** Binding energies per nucleon $E/A$ for different nuclei. Theoretical values correspond to the unperturbed cases ($\lambda_i = 0$). Calculations were performed with NNLO$_{sat}$ [30] interaction, $\hbar\omega$=20 MeV, and model space specified by $N_{max}$ in the legend. Black rectangles represent the experimental values taken from Ref. [52].

The SCGF calculations were performed using a basis of spherical harmonic oscillator wave functions with oscillator energy $\hbar\omega$=20 MeV. The model space was limited to states with principal quantum numbers smaller or equal than $N_{max}$. In terms of computational resources, even if the present technology allows calculations up to $N_{max} = 13$, this would require a too large amount of CPU hours to complete the full set of perturbations required in the current analysis. We thus decided to limit the model space to $N_{max} = 9$. This model space may converge poorly to the experimental values, but, in the first attempt at this approach, our attention focus on the validity of the method rather than on the accurate prediction of the experimental masses.

In Fig. 2, we show binding energies per nucleon, $E/A$, for seven nuclei $^{16}$O, $^{24}$O, $^{34}$Si, $^{36}$S, $^{40}$Ca, $^{48}$Ca, and $^{56}$Ni. Each nucleus is in the configuration corresponding to fully filled spherical orbitals. These nuclei represent all available convergent ADC(3) calculations among the magic systems between $^{16}$O and $^{56}$Ni.

Values of $E/A$ are reported for three choices of the size of model space: $N_{max} = 7$,

9, and 11, to be compared with the experimental values taken from Ref. [52]. The convergence in terms of $N_{max}$ is not fully achieved; nevertheless, with increasing $N_{max}$, most binding energies decrease towards the experimental values. The model space of $N_{max} = 7$ appears to be too small to reproduce experimental results. For $N_{max} = 11$, the maximum difference with experiment appears for $^{56}$Ni, with the difference of around 3% of the total energy. This discrepancy is a combination of the 1% error on the correlation energy that is associated with the ADC(3) truncation [37] and of the accuracy of state-of-the-art chiral nuclear forces [53, 51]. In absolute terms, the calculated binding energy of $^{56}$Ni is more than 10 MeV above the measured value. Such a deviation is much larger than the standard deviation of typical EDFs, which is of the order of 1 MeV [24, 54]. Clearly, with the model space truncation discussed below, the overall accuracy with respect to predicting the experiment is too poor to obtain novel functionals capable of reducing the discrepancies between the EDF approach and the experiment. However, our goal is to reproduce the *ab initio* energies, irrespective of their detailed agreement with experiment.

Typical *ab initio* computations subtract the kinetic energy of the center of mass to directly access the intrinsic ground state energy [55]. This implies adding a one- and a two- body correction terms to the nuclear Hamiltonian [56, 37]. The result of retaining only the one-body term is shown in Fig. 2 for $N_{max} = 9$ as "1bodycm" and it amounts to an overcorrection. However, this additional discrepancy will not affect our goal of investigating the consistence between microscopic results and the functionals that they generate. For our purposes, it is technically simpler to drop the two-body center-of-mass correction because it avoids further approximations in the way the kinetic energy is treated in the SCGF and EDF approaches. Since the EDF results also depend on the number of oscillator shells included in the model space [57], we fixed the model spaces of both approaches to be $N_{max} = 9$ (1bodycm) for the following analysis.

In the model functional, Eq. (8), we consider generators of zero-range contact interaction (Skyrme-like) defined in Eq. (10), namely, $\hat{V}_T^\rho$, $\hat{V}_T^{\Delta\rho}$, $\hat{V}_T^\tau$, $\hat{V}_T^{J1}$, $\hat{V}_{W0}$, and $\hat{V}_{t3}$. Subscripts $T = 0$ and 1 denote generators inducing terms of the functional that depend on isoscalar and isovector densities, respectively. Such a link between generators and densities is valid only on the EDF level; for the chiral interactions used in our SCGF computations, already at the next-to-leading order in powers of the interaction, expectation values of generators contain contributions from both isoscalar and isovector densities.

We perturbed ground states of each of the seven selected nuclei using four different intensities of the perturbation strength $\lambda_i$, separately for each of the 10 generators. In this way, we obtained 284 converged results‡, which represent our full database of the perturbed and unperturbed ground-state energies. The choice of using non-zero values of $\lambda_i$ separately for each $\hat{V}_i^{gen}$ represents a compromise between the volume of calculations and a coverage of the full manifold in the space of perturbations $\lambda_i$. Obviously, the larger the $\lambda_i$'s the wider is the density space probed, however, if perturbations are too strong, the numerical SCGF solutions may diverge. In practice, for each generator, we have found a most suitable range of values of $\lambda_i$'s that were used in the final calculations.

___________

‡ For $^{36}$S, 3 data points did not converge.

### 3.1. Estimated errors on the SCGF calculations

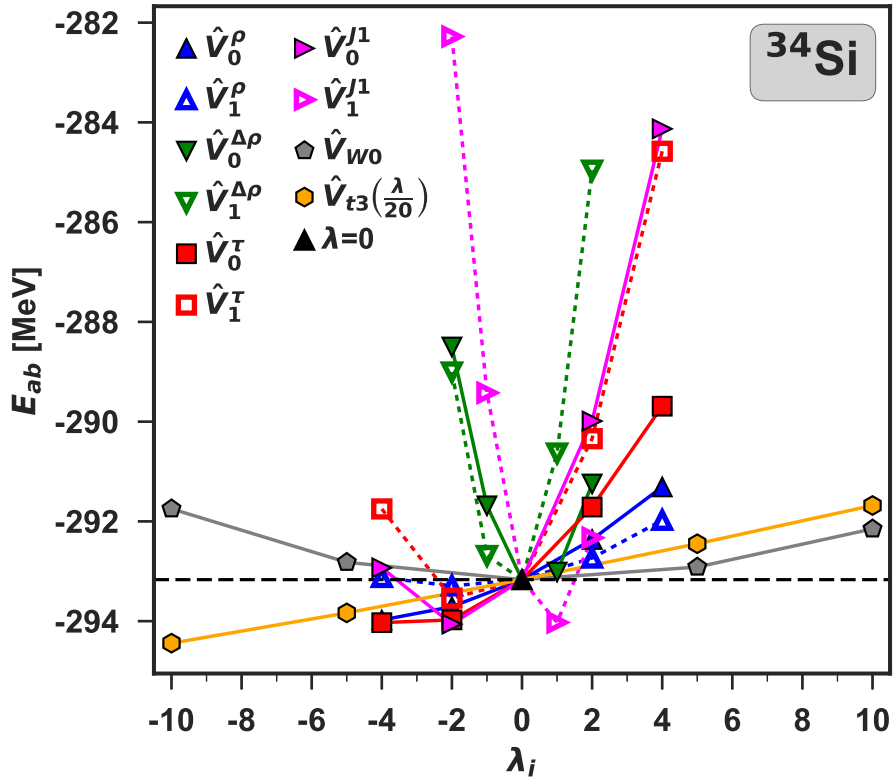

**Figure 3.** Perturbed *ab initio* energies compared with the unperturbed energy in $^{34}$Si. Symbols shown in the legend correspond to different generators $\hat{V}_i^{\text{gen}}$. The full triangle represents the unperturbed energy at $\lambda_i = 0$ and the dashed line shows the reference value of $E^{\text{ab}}(0)$.

Figure 3 shows the *ab initio* energies in function of $\lambda_i$ calculated for different perturbations $\hat{V}_i^{\text{gen}}$ in $^{34}$Si. We had expected that the value at $\lambda = 0$ (black triangle) would be the one with the lowest energy $E^{\text{ab}}(0)$, because for any state different than the exact ground state $|\Psi(0)\rangle$, the variational principle stipulates that

$$E^{\text{ab}}(\lambda_i) \geq E^{\text{ab}}(0), \tag{27}$$

assuming, of course, that the *ab initio* energies are calculated exactly. From the plot, it is evident that there are cases of energies $E^{\text{ab}}(\lambda_i)$ smaller than $E^{\text{ab}}(0)$ in violation of the variational principle. For the perturbations induced by the three-body generator $\hat{V}_{t3}$ (hexagons), the energy does not present a minimum, but increases monotonically with $\lambda_i$. This effect can be partially related to the way the contribution of the three-body interaction is extracted. In fact, we can estimate only the leading order as in Eq. (7).

The violation of the variational principle must be traced back to a series of approximations used to calculate the SCGF results. To make any sensible use of these metadata when fitting the functional coupling constants, we thus have to estimate the uncertainties associated with the SCGF calculations.

The major source of error in our calculation is probably related to the subtraction procedure of the perturbation energies from solutions given by the minimisation of the Routhian. For simplicity, we further assume that only one perturbing potential contributes to this uncertainty. The resulting subtraction error, $\Delta E_{\mathrm{S}}^{\mathrm{ab}}(\lambda_i)$, associated with the value of $E^{\mathrm{ab}}(\lambda_i)$ can be estimated as

$$\Delta E_{\mathrm{S}}^{\mathrm{ab}}(\lambda_i) = \left| \langle \lambda_i \hat{V}_i^{\mathrm{gen}}(\lambda_i) \rangle_{\mathrm{RL}} - \langle \lambda_i \hat{V}_i^{\mathrm{gen}}(\lambda_i) \rangle_{\mathrm{NLO}} \right|, \tag{28}$$

where $\langle \rangle_{\mathrm{RL}}$, $\langle \rangle_{\mathrm{NLO}}$, indicates the truncation respectively up to ring and ladder RPA, to NLO approximation. Such error can be viewed as a relative error between the perturbed solutions and the unperturbed one. In Figure 4, we show the subtraction errors $\Delta E_{\mathrm{S}}^{\mathrm{ab}}(\lambda_i)$ calculated for perturbations induced by the potential $\hat{V}_0^{J1}$ in $^{34}$Si. As expected, this error is zero for $\lambda_i = 0$ and grows rapidly with increasing values of the perturbation strength parameter $\lambda_i$.

We also found another way to estimate uncertainties of the SCGF calculations, which is independent of the approximately calculated average values of the perturbation potentials $\hat{V}_i^{\mathrm{gen}}$. Indeed, we can determine such estimates by using the well-known Hellmann-Feynman theorem [58]. For any Hamiltonian $\hat{H}_\lambda = \hat{H}_0 + \lambda \hat{V}^{\mathrm{pert}}$ that depends explicitly on the parameter $\lambda$, the theorem states that

$$\frac{\mathrm{d}E_\lambda}{\mathrm{d}\lambda} \equiv \frac{\mathrm{d}}{\mathrm{d}\lambda} \langle \Psi(\lambda) | \hat{H}_\lambda | \Psi(\lambda) \rangle = \langle \Psi(\lambda) | \frac{\partial \hat{H}_\lambda}{\partial \lambda} | \Psi(\lambda) \rangle = \langle \Psi(\lambda) | \hat{V}^{\mathrm{pert}} | \Psi(\lambda) \rangle. \tag{29}$$

Eq. (29) is valid under the condition that $|\Psi(\lambda)\rangle$ is an eigenstate of $\hat{H}_\lambda$, $\hat{H}_\lambda | \Psi(\lambda) \rangle = E_\lambda | \Psi(\lambda) \rangle$, or an Hartree-Fock wave function [59], or a variational wave function [60]. However, when the wave function is expanded in a truncated basis [61], or it is a solution of a perturbative expansion [60], the Hellman-Feynman theorem is violated. The ground-state wave function in the SCGF method is not variational because the ADC(3) approximation is a truncated expansion. The degree to which the Hellman-Feynman theorem is violated then illustrates the degree of violation of the variational principle.

The derivative at $\lambda = 0$ in the left-hand side of Eq. (29) can be, for small values of $\lambda$, determined by the finite difference method,

$$\frac{\mathrm{d}E_\lambda}{\mathrm{d}\lambda}(0) \approx \frac{E(\lambda) - E(-\lambda)}{2\lambda}. \tag{30}$$

In our numerical test, we studied the case of the perturbation given by $\hat{V}^{\mathrm{pert}} = \hat{V}^{\mathrm{ab}} + \hat{V}^{\mathrm{Coul}}$, that is, by the full potential $V^{\mathrm{tot}}(0)$, Eq. (24), that defines $\hat{H}^{\mathrm{ab}}$ at $\lambda_i = 0$.

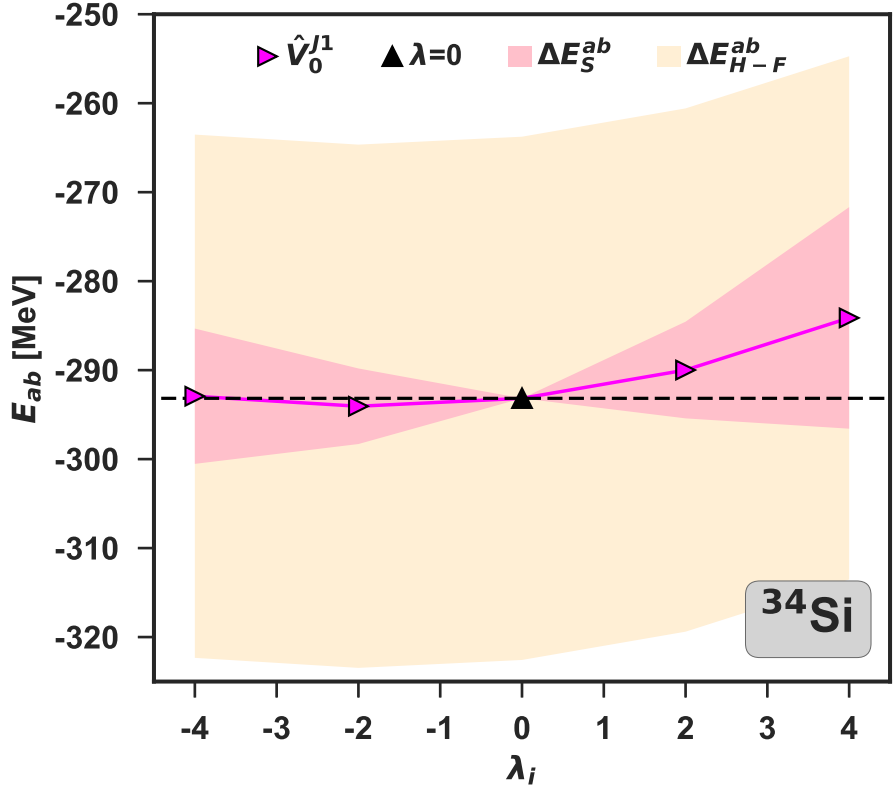

**Figure 4.** Examples of errors associated with the *ab initio* energies $E^{\mathrm{ab}}(\lambda_i)$, estimated in $^{34}$Si for perturbations related to generator $\hat{V}_0^{J1}$. The darker shadow represents the error $\Delta E_{\mathrm{S}}^{\mathrm{ab}}$ given by Eq. (28). The lighter shadow represents the error $\Delta E_{\mathrm{H\text{-}F}}^{\mathrm{ab}}$ extracted from the Hellmann-Feynman theorem, Eq. (31).

This compares the finite difference of the energy calculated for the perturbed cases ($\lambda$ and $-\lambda$) with the average value of the interaction energy in the unperturbed case ($\lambda = 0$). Such a comparison offers an estimate of the difference between the approximated energy in the ADC(3) method and the exact energy. Consequently, we define the error of the *ab initio* energy as

$$\Delta E_{\mathrm{H\text{-}F}}^{\mathrm{ab}} = \left| \frac{\mathrm{d}E_\lambda}{\mathrm{d}\lambda}(0) - \langle \Psi(0) | \hat{V}^{\mathrm{tot}}(0) | \Psi(0) \rangle \right|, \tag{31}$$

where subscript H-F stands for the error extracted from the Hellmann-Feynman theorem. This error represents an absolute error of the total energy that is due to the approximated solution of the SCGF method. It depends on the nucleus, but we assume it is independent of the perturbation, namely the value calculated at $\lambda=0$ is attributed to all perturbed and unperturbed total energies of a given nucleus.

In the case of $N_{\mathrm{max}}= 9$ (1bodycm), the violation of the Hellmann-Feynman theorem, as defined in Eq. (31), is for lighter (heavier) nuclei studied in this paper equal to about 1% (3-4%) of the total energy. This error is larger than the estimate provided in Ref. [37], because it gives a cumulative effect resulting from the ADC(3) truncation and reduced model space.

In Figure 4, we also show the Hellmann-Feynman error $\Delta E_{\text{H-F}}^{\text{ab}}$ determined in $^{34}$Si. We see that for this particular generator, $\hat{V}_0^{J1}$, $\Delta E_{\text{H-F}}^{\text{ab}}$ is more than twice larger than $\Delta E_{\text{S}}^{\text{ab}}$. We checked that in most cases considered in this paper $\Delta E_{\text{H-F}}^{\text{ab}} > \Delta E_{\text{S}}^{\text{ab}}$, even in very light nuclei. Given the very different magnitude and sources of the subtraction and Hellmann-Feynman errors, we decided to keep them separate and perform two independent analyses of the coupling constants.

In view of the identified uncertainties, we can now conclude that the explicit violation of the variational principle obtained for $E^{\text{ab}}(\lambda_i \neq 0)$, see Fig. 3, can be considered acceptable consequences of the inherent imprecision encountered in the SCGF approach.

### 3.2. Linear regression analysis to determine the coupling constants

From the set of *ab initio* calculations described in Sec. 2.3, we obtained $d = 284$ equations (25) whose left-hand sides represent regression dependent variable $y_i$, $i = 1 - 284$, and whose $p$ expectation values on the right-hand sides form the matrix of features, that is,

$$y_i = \sum_{j=1}^{p} \mathcal{J}_{ij} C_j, \tag{32}$$

where each coupling constant $C_j$ corresponds to a generator of the model functional given in Eq. (8).

Introducing for compactness the vector notation: $\boldsymbol{C} = (C_1, \ldots, C_p)$ and $\boldsymbol{y} = (y_1, \ldots, y_d)$, we build the penalty function by of the least-square method as

$$\chi^2(\boldsymbol{C}) = \frac{1}{d-p} \left( \mathcal{J}\boldsymbol{C} - \boldsymbol{y} \right)^T W \left( \mathcal{J}\boldsymbol{C} - \boldsymbol{y} \right), \tag{33}$$

where the weight matrix $W$ is a diagonal matrix with elements $W_{ii} = w_i$. We define the weight of each data point as the inverse of the estimated error squared,

$$w_i = \frac{1}{(\Delta y_i)^2}, \tag{34}$$

where $\Delta y_i$ is composed of three contributions [62],

$$(\Delta y_i)^2 = (\Delta y_i^{\text{ab}})^2 + (\Delta y^{\text{num}})^2 + (\Delta y^{\text{mod}})^2. \tag{35}$$

$\Delta y_i^{\text{ab}}$ is the error attributed to the SCGF approach, namely, we take it as $\Delta E_{\text{S}}^{\text{ab}}$ (28) or $\Delta E_{\text{H-F}}^{\text{ab}}$ (31). $\Delta y^{\text{num}}$ is the numerical precision of the SCGF calculations, which is smaller that $5 \times 10^{-5}$ MeV and can be neglected. $\Delta y^{\text{mod}}$ represents the error associated with the model itself, which is entirely unknown, and thus it has to be tuned to normalise the penalty function $\chi^2$. Starting from an arbitrary value, $\Delta y^{\text{mod}}$ can be increased iteratively up to the value at which the $\chi^2$ approaches the value of 1. Then, the penalty function in Eq. (33) satisfies the typical statistical normalisation condition $\chi^2(\boldsymbol{C}) \to 1$ at the minimum, and $\Delta y^{\mathrm{mod}}$ acquires interpretation of the Birge factor [63].

Minimisation of $\chi^2$ gives the solution $\boldsymbol{C}_{min}$, covariance matrix $\mathcal{K}$, statistical error associated with the parameters, $\Delta \boldsymbol{C}_{min}$, and propagated errors of observables [64, 62, 65].

### 3.3. Fitted coupling constants

We minimise the $\chi^2$ defined in Eq. (33) using the two types of errors discussed in previous section. We thus obtain two sets of results: the one labeled $\mathrm{D_S}(10)$, that is, obtained using the errors defined in Eq. (28) and that labeled $\mathrm{D_{H\text{-}F}}(10)$ – using the errors defined in Eq. (31).

Given the large uncertainties of the metadata and the fact that they could be obtained only in fairly light nuclei with small isospin asymmetry, the resulting coupling constants are poorly determined with quite large error bars. This is particularly evident for the $\mathrm{D_{H\text{-}F}}(10)$ set of parameters. The relative errors of the isovector coupling constants are often of the order of 100%, meaning that the obtained values are compatible with 0. The very large errors mean that the $\chi^2$ surface is fairly flat in these particular directions of the parameter space. As a consequence, when used to calculate atomic nuclei, the obtained coupling constants often immediately lead to finite-size instabilities [66].

The quality of the fit can be easily judged by inspecting relative residuals of Eq. (25), defined as

$$\mathrm{Residuals} \equiv \frac{\sum_j C_j \, V_j^{\mathrm{gen}}[\rho^{\mathrm{ab}}(\lambda_i)] - V^{\mathrm{ab}}(\lambda_i)}{V^{\mathrm{ab}}(\lambda_i)}, \qquad (36)$$

and for $^{56}$Ni shown in Fig. 5(a) for $\mathrm{D_S}(10)$ and 5(b) for $\mathrm{D_{H\text{-}F}}(10)$. For a good-quality fit, we would expect the residuals lie close to the dashed horizontal line, which indicates zero values. Rapid departures of residuals from zero, especially for the second-order isovector coupling constants, illustrate the fact that a reasonable description of the *ab initio* results in terms of Skyrme functional generators could not be obtained.

For all nuclei and generators studied here, the residuals corresponding to the unperturbed systems ($\lambda=0$) are around zero, whereas when $\lambda_i$ moves away from zero, the residuals increase. In addition, they are not normally distributed, but they exhibit clear trends, meaning that the hypothesis formulated in Eq. (25) is not correct. Shadows shown in Fig. (5)(a) represent the propagated errors of the corresponding observables obtained in the fit $\mathrm{D_S}(10)$. For $\mathrm{D_{H\text{-}F}}(10)$, the propagated errors are much larger, so in Fig. (5)(b), for clarity we only show the one corresponding to generator $\hat{V}_{W0}$. We note that the propagated errors related to the violation of the Hellmann Feynman theorem are much larger than those corresponding to the subtraction errors, and thus they make it very hard to draw reasonable conclusions about the structure of the residuals.

We tested the performance of the obtained coupling constants in the description of infinite nuclear matter. For $\mathrm{D_S}(10)$, we obtain a fairly reasonable saturation properties of infinite nuclear matter, with an energy per particle $E/A = -14.6 \pm 0.2$ MeV and a value for the saturation density of $\rho_0 = 0.132 \pm 0.002$ fm$^{-3}$. The D$_{\text{H-F}}$(10) provides slightly different values for the binding energy $E/A = -16.0 \pm 1.2$ MeV and saturation density $\rho_0 = 0.128 \pm 0.01$ fm$^{-3}$, but with much larger error bars than in D$_{\text{S}}$(10). For both sets, the nuclear incompressibility $K$ is also in the range of acceptable values [67]; in particular, we have $K = 380 \pm 17$ MeV for D$_{\text{S}}$(10) and $K = 393 \pm 143$ MeV for D$_{\text{H-F}}$(10). Other properties of infinite matter, such as the symmetry energy and its first derivative, are poorly determined, probably because the set of nuclei used for the fit is not rich enough to describe isovector properties of the nuclear medium sufficiently well. In particular, we find that the symmetry energy has a value compatible with zero within the error bars.

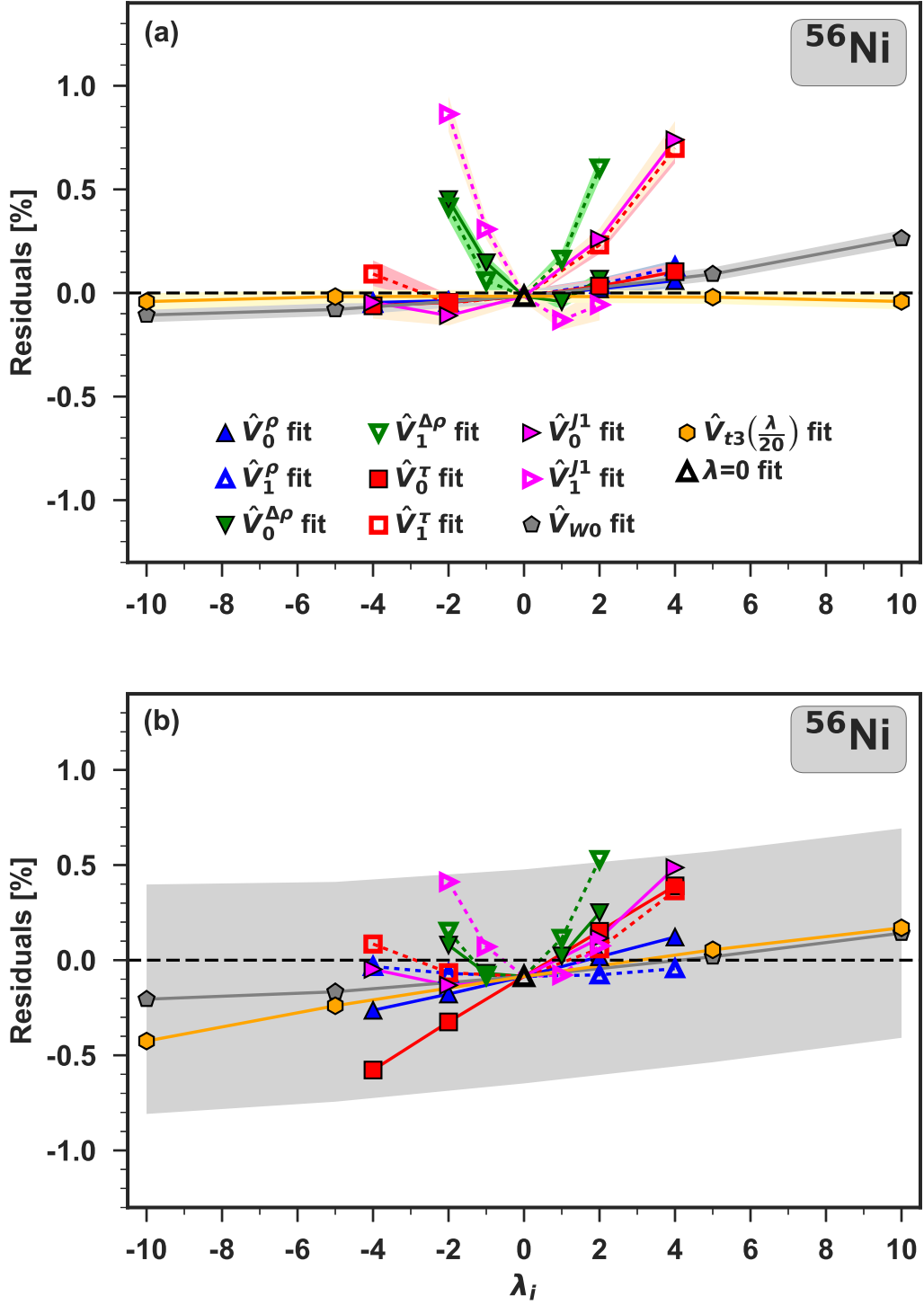

**Figure 5.** Residuals calculated for $^{56}$Ni with the parametrizations $D_S(10)$ (a) and $D_{H\text{-}F}(10)$ (b). Shadows show the corresponding propagated errors. For clarity, in (b) we only show the one corresponding to generator $\hat{V}_{W0}$.

Another major drawback of the derived coupling constants is an unrealistic value of the effective mass. Although the effective mass is not strictly speaking an observable, we can extract information about its value from other many-body methods [68]. For the isoscalar effective mass, an acceptable range of values is $m^*/m \in [0.7 - 0.9]$, although it is worth mentioning that also other values are found in the literature. Both sets of coupling constants, D$_{\text{S}}$(10) and D$_{\text{H-F}}$(10), give effective masses that are off by roughly an order of magnitude (cf. Fig. 7), and this is probably the main reason why they do not lead to realistic results when used in calculations of finite nuclei.

### 3.4. Constraints on the nuclear matter properties

Since a simple least-square minimisation does not provide us with satisfactory values of the effective mass, we also performed a constrained linear regression to drag the coupling constants toward reasonable values of $m^*/m$. In this way, we want to test if the poor determination of the coupling constants, reflected in their large errors, can be exploited to improve values of the effective mass. Linear regression with constraints is a procedure in the spirit of the Bayesian inference, where a prior information about the parameters is known. Here we consider it in the form of the Tikhonov regularisation [69] or ridge regression [70], which consists of minimising a penalty function of the form

$$\chi_{\text{T}}^2(\boldsymbol{C}) = \frac{1}{d-p+f} \left[ (\boldsymbol{y} - \mathcal{J}\boldsymbol{C})^T W (\boldsymbol{y} - \mathcal{J}\boldsymbol{C}) + \lambda_{\text{T}}(\boldsymbol{b} - \boldsymbol{Q}[\boldsymbol{C}])^T (\boldsymbol{b} - \boldsymbol{Q}[\boldsymbol{C}]) \right]. \quad (37)$$

The Tikhonov parameter $\lambda_{\text{T}}$ is a real positive number and $\boldsymbol{b} = \boldsymbol{Q}[\boldsymbol{C}]$ represents a system of $f$ linear equations in parameters $\boldsymbol{C}$ with constant terms $\boldsymbol{b}$. The final values of $\boldsymbol{C}$ will depend on $\lambda_{\text{T}}$, which is unique for the $f$ independent constraining equations. Eq. (37) is defined as a sum of two terms: $\chi_{\text{data}}^2$, which depends on weights $W$, and $\chi_{\text{constr}}^2$, which depends on $\lambda_{\text{T}}$. Increasing the value of the Tikhonov parameter gives more importance to $\chi_{\text{constr}}^2$ and eventually may deteriorate the description of data represented by $\chi_{\text{data}}^2$.

We use one constraint, namely the definition of the in-medium isoscalar effective mass,

$$\frac{m^*}{m}(\rho_{\text{sat}}) = \left[ 1 + \frac{2m}{\hbar^2} C_0^\tau \rho_{\text{sat}} \right]^{-1}. \quad (38)$$

It is important to note that $\frac{m^*}{m}$ depends explicitly only on one coupling constant, $C_0^\tau$, however, implicitly, through the saturation density $\rho_{\text{sat}}$ it non-linearly depends on all other coupling constants. We set the target value of the effective mass to $b_1 \equiv m^*/m = 0.70$ and in the Tikhonov term we varied the value of $\log_{10}\lambda_T$ from -4 to 2. As an example, in $\chi^2_{\text{data}}$ we took the inputs used for $D_{\text{H-F}}(10)$.

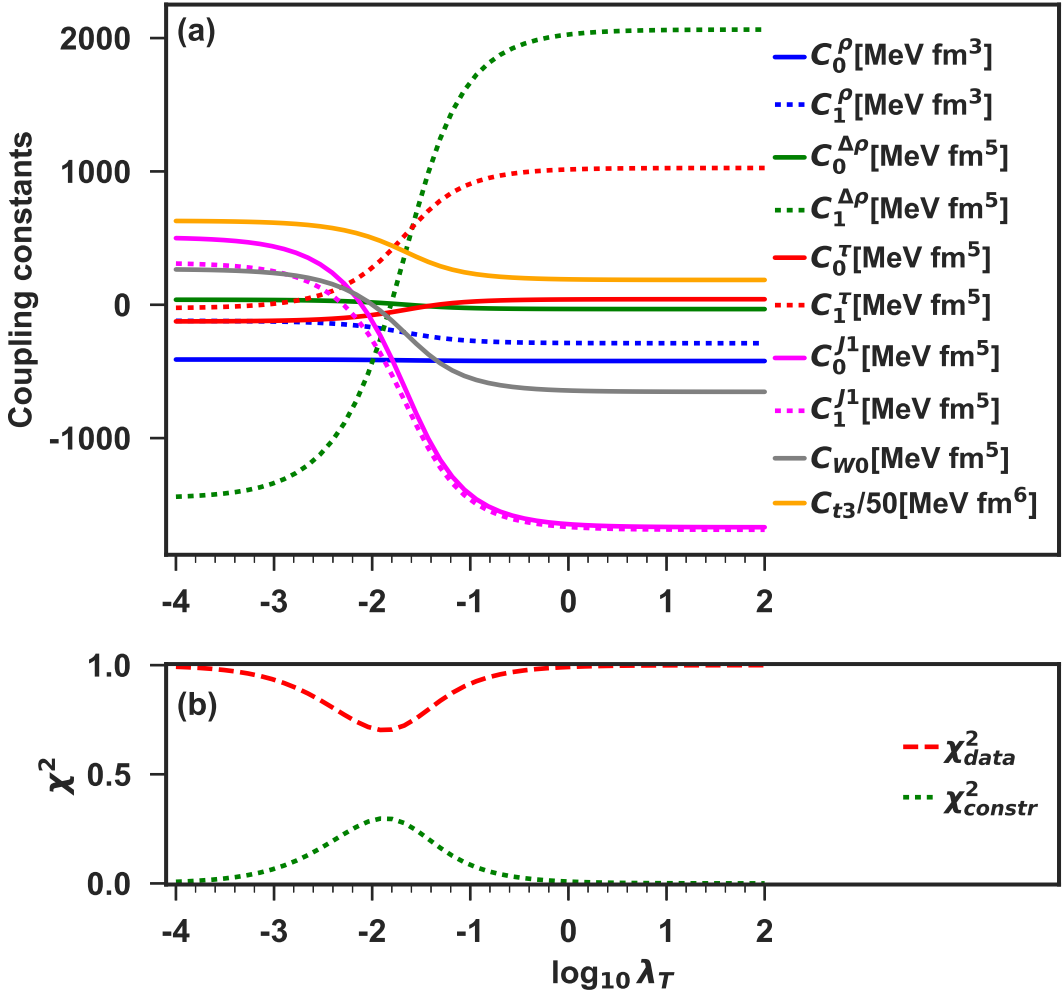

**Figure 6.** (a) Results obtained for the $D_{\text{H-F}}(10)$ coupling constants with $m^*/m$ constrained to 0.70 using the Tikhonov regularisation. (b) Contributions of the data, $\chi^2_{\text{data}}$, and constraint, $\chi^2_{\text{constr}}$, to the total penalty function $\chi^2_T$ (37).

In Fig. 6(a), we show the obtained evolution of the constrained parameters $\boldsymbol{C}$ in function of the Tikhonov parameter. As one can see, changes in the parameters happen around $\log_{10}\lambda_T \approx -2$. Beyond this region, their values are quite stable. $C_1^{\Delta\rho}$, $C_0^{J1}$, $C_1^{J1}$ and $C_{t3}$ are the coupling constants to which the constraint makes the largest impact. We already noted that these coupling constants were poorly determined by the unconstrained regression. In Fig. 6(b), we show separate contributions from the data points, $\chi^2_{\text{data}}$, and from the constraint, $\chi^2_{\text{constr}}$, (see Eq. (37)). We see that the contribution of the constraint increases with $\lambda_{\text{T}}$ up to $\log_{10}\lambda_{\text{T}} \approx$ -1.8 and after that it quickly drops to zero. At the peak of $\chi^2_{\text{constr}}$, the coupling constants begin to adjust to the requested value of $m^*/m$. A further increase of $\lambda_{\text{T}}$ decreases $\chi^2_{\text{constr}}$, because $b_1 - Q_1[\boldsymbol{C}] \to 0$, whereas the reproduction of the data points deteriorates and the differences $\boldsymbol{y} - \mathcal{J}\boldsymbol{C}$ increase.

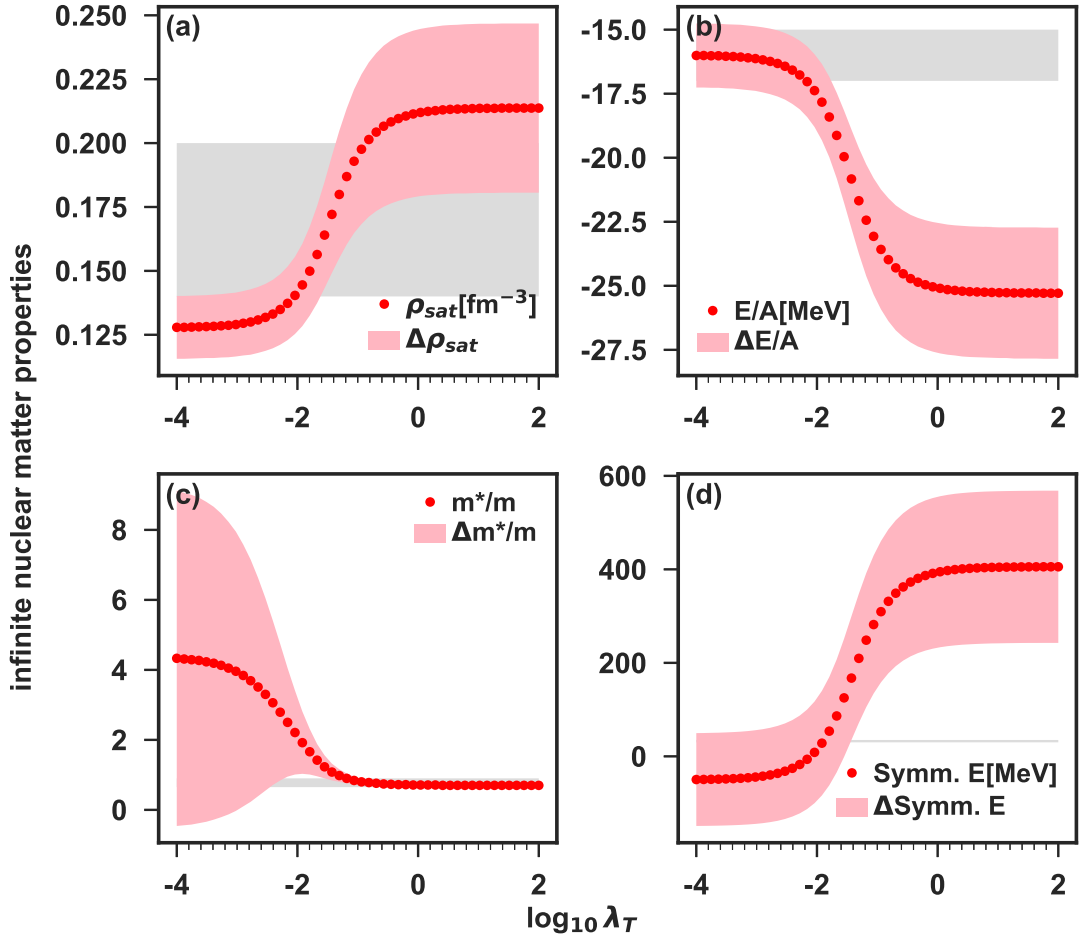

**Figure 7.** Infinite nuclear matter properties: $\rho_{\text{sat}}$ (a), $E/A$ (b), $m^*/m$ (c), and symmetry energy $J$ (d). obtained by the Tikhonov regularisation of the $D_{\text{H-F}}(10)$ coupling constants with $m^*/m$ constrained to 0.70. Shadows show the propagated error bars and grey regions indicate the ranges of empirical values.

In Fig. 7, we present nuclear matter properties $\rho_{\text{sat}}$, $E/A$, $m^*/m$, and symmetry energy $J$ obtained by the Tikhonov regularisation of the $D_{\text{H-F}}(10)$ coupling constants with $m^*/m$ constrained to 0.70. For small $\lambda_{\text{T}}$, the obtained values are equal to those corresponding to the original $D_{\text{H-F}}(10)$ results. In the region of $\log_{10}\lambda_{\text{T}}$ between -2 and 0, nuclear-matter properties change abruptly, and effective mass is dragged towards the target value already at $\log_{10}\lambda_{\text{T}} \simeq -1$. Beyond $\log_{10}\lambda_{\text{T}} \simeq 0$ nuclear-matter properties do not vary much. The curves for $\rho_{\text{sat}}$, $E/A$, and symmetry energy cross their regions of empirical values (shown in grey) before setting to values far away from the standard nuclear-matter values. Such crossing occurs at slightly different values of $\lambda_T$ for the three quantities and before the effective mass reaches its empirical range. When the effective mass approaches the target value of 0.70, the energy per particle $E/A$ becomes too low and the symmetry energy becomes very large. We conclude that there is no region of the Tikhonov parameter where all four nuclear-matter quantities would be in their expected domains, even when the propagated errors (represented by shaded areas) are taken into account. A possible reason for this failure can probably be traced back to strong correlations between the Skyrme nuclear-matter properties, cf. Ref. [71].

## 4. Conclusions

Applying the methodology suggested in Ref. [28], we studied the link between the nuclear Skyrme functional and the NNLO$_{\text{sat}}$ chiral interaction used within the *ab initio* Self-Consistent Green's Function calculations in ADC(3) approximation. We performed the *ab initio* calculations in seven light closed shell nuclei by perturbing their ground states with ten functional generators that define Skyrme functional. By employing the linear regression method, the obtained metadata were used to derive the functional coupling constants. We analysed two possible sources of uncertainties of the *ab initio* calculations: the first one related to approximate determination of average values of two-body potentials and the second one to an imprecise determination of nuclear ground states, for which we employed the Hellmann-Feynman theorem.

The obtained values of Skyrme coupling constants were very different than those typically obtained in phenomenological adjustments to nuclear observables. We have identified several possible reasons of such a result: First, it appears that a relatively high level of uncertainties arising in the *ab initio* calculations induces large uncertainties of the derived coupling constants, which then propagate to large uncertainties of the nuclear matter properties and to instabilities when solving the self-consistent equations. Second, it appears that the *ab initio* energies are poorly reproduced by the terms in the functional generated by the Skyrme zero-range potentials. Third, it appears that the information contents within the perturbed ground-state energies of light semi-magic nuclei is insufficient to properly determine Skyrme coupling constants, especially those corresponding to second-order terms depending on isovector densities.

Certainly, future research may be focused on applications of *ab initio* technologies with improved overall precision, which would better correspond to the ambition of reducing discrepancies between the phenomenological EDF results and experiment. Another promising avenue would be to repeat present analyses by using finite-range functional generators. However, the most challenging aspect of the methodology proposed in Ref. [28] is the fact that it is based, similarly as most other generic *ab initio* DFT approaches, on the Levy-Lieb construction that essentially pertains to variational studies of ground-state energies. This is in opposition to the methodology of adjusting functionals directly to experimental data, where one uses not only ground-state energies, but also other essential observables like radii, deformations, or transition probabilities.

## Acknowledgement

This work was supported in part by the Academy of Finland and University of Jyväskylä within the FIDIPRO program, by the York STFC Grants No. ST/M006433/1 and No. ST/P003885/1, by the Surrey's STFC grants No. ST/P005314/1 and No. ST/L005816/1, and by the Polish National Science Centre under Contract No. 2018/31/B/ST2/02220. Numerical calculations were performed using computational resources provided by the CSC-IT Center for Science, Finland, and using the DiRAC Data Analytic system at the University of Cambridge, operated by the University of Cambridge High Performance Computing Service on behalf of the STFC DiRAC HPC Facility (www.dirac.ac.uk). This equipment was funded by BIS National E-infrastructure capital grant (ST/K001590/1), STFC capital grants ST/H008861/1 and ST/H00887X/1, and STFC DiRAC Operations grant ST/K00333X/1. DiRAC is part of the National e-Infrastructure.